\documentstyle[prl,epsfig,twocolumn,aps]{revtex}
\begin{document}
\twocolumn[\hsize\textwidth\columnwidth\hsize\csname@twocolumnfalse%
\endcsname
\title{Mounding Instability and Incoherernt Surface Kinetics 
 }  
\author{S.V. Ghaisas} 
\address{
 Department of Electronic Science, University of Pune, Pune 411007,
India}

\date{\today}
                  
\maketitle 

\begin{abstract}  
 Mounding instability in a conserved growth from vapor is analysed 
within the framework of adatom kinetics on the growing surface. The 
analysis shows that depending on the local structure on the surface, 
kinetics of adatoms may vary, leading to disjoint regions in the sense 
of a continuum description. This is manifested particularly under the 
conditions of instability. Mounds grow on these disjoint regions and 
their lateral growth is governed by the flux of adatoms hopping across 
the steps in the downward direction. Asymptotically $\ln t$ dependence 
is expected in 1+1- dimensions. Simulation results confirm the prediction. 
Growth in 2+1- dimensions is also discussed.  
   
\end{abstract} 

] 
\narrowtext
Mounding instability was experimentally observed and proposed by 
Johnson {\it et. al}\cite{len} during growth of GaAs on (001) GaAs 
substrate.Initially, activation difference (Schwoebel - Ehrlich (SE) barrier
)\cite{se} between adatoms hopping on the plane and the one crossing the 
step edge was considered responsible \cite{vill1}. Later it was shown 
that edge diffusion can also lead to similar effect \cite{tl}. One of the 
issues related to growth of mounds has been the temporal dependence of 
mound growth. Based on various forms of continuum equations the lateral 
growth is expected to have a time dependence $\sim t^{s}$ where $s$ takes 
values from 0.0 to 1/4 \cite{pl}. Similarly, the width of the 
interface is predicted to follow the power law $t^{\beta}$ with $\beta$ 
varying from 1/3 onwards \cite{pl}. All these predictions are based 
on the assupmtion that the underlying conserved growth equation 
describing non equilibrium growth, is valid over the entire substrate. 
In the following we show that under the conditions of instability and  
low temperatures, this assumption fails. Consequently the growth of mounds 
is governed laterally by the adatom kinetics across the mound boundaries. 
We show this by developing growth equation over a stepped surface in 1+1 
dimensions using kinetics of adatoms and steps. This helps establish the 
correspondence between kinetic processes and terms in the growth 
equation. We assume that only mechanism of relaxation is by diffusion 
of adatoms. This allows identification of process and corresponding term 
uniquely. These assumptions are expected to be valid at low temperature, 
where evaporation is negligible.  
Once the kinetic processes 
leading to various terms in growth equation are identified, presence 
or absence of such terms in various regions on the surface can be predicted. 
This allows us to classify different regions on the surface according 
to the growth equation followed there. 

Consider growth on a one dimensional substrate. Fig.1 shows the stepped 
region under consideration. Growth proceeds through randomly falling 
adatoms on the surface that relax by diffusing on the stepped terraces. 
Adatoms with zero nearest neighbors (nn) 
are mobile while those with more than zero 
nn will have negligible mobility. Further, desorption and dissociation 
from the steps is also negligible at low temperature. Under the conserved 
growth conditions it is possible to write formally the growth equation 
in the form $\partial_{t} h({\bf x},t)=\nabla \cdot{\bf j}({\bf x},t) 
+F$, where, $F$ is incident flux, $h({\bf x},t)$ is height function 
and ${\bf j}({\bf x},t)$ is particle current. An uphill current on 
a tilted substrate indicates instability while downhill indicates stable 
Edward-Wilkinson (EW)\cite{ew} type growth\cite{kr1}.   
Let $l_{d}$ be the 
average length travelled by an adatom before getting attached to another 
adatom or step. The density of steps can be expressed as $\frac{|m|}{1+|m|}$
. Let $P_{A}$ and $P_{B}$ be the relative probabilities for hopping 
across the sites A and B in Fig. 1. By considering current due to the 
downward hops and that due to the in-plane hops seperately, one can show that 
the resultant nonequilibrium current is given by \cite{prep}      
\begin{equation}  
   {\bf j}_{s}=\frac {\hat n |m|F(P_{B}-P_{A})
}{2(1+|m|)(l^{-1}_{d}+|m|a^{-1})}
\end{equation} 
Where $\hat n$ denotes +ve x direction.  
Presence of $l^{-1}_{d}$ in the denominator accounts for the nucleation 
effect on larger terraces. In this expression, local terrace width is 
$(l^{-1}_{d}+a^{-1}|m|)^{-1}$. However due to the relative velocity between 
two adjacent terraces, the local terrace width changes. The velocity 
difference will be proportional to the $\frac{\partial j(x,t)}{\partial x}$. 
Including this dynamical effect, the expression for the current becomes,   
\begin{eqnarray}  
  {\bf j}(x)&=&\frac {\hat n |m|F (P_{B}-P_{A})}{2(1+|m|)
(l^{-1}_{c}+|m|a^{-1})}
             \nonumber\\
&&-\frac {\hat n F}{4} 
\partial_{x}
\left(\frac {|m|}{(1+|m|)(l^{-1}_{c}+|m|a^{-1})}\right)^{2}    
\end{eqnarray}

Next, we argue that every downward hop introduces height-height correlation, 
hence will give rise to all the stabilizing terms in a growth equation. 
Under the tilt independent current conditions, the lowest of such terms 
is $K\frac{\partial^{3} h}{\partial x^{3}}$. Thus the current 
on the stepped surface will be,  
\begin {eqnarray}
  {\bf j}(x)&=&\frac {\hat n |m|F(P_{B}-P_{A})}{2(1+|m|)(l^{-1}_{c}+
|m|a^{-1})}\nonumber \\ 
&&-\frac {\hat n|m|F}{4}   
\partial_{x}\left(\frac {|m|}{(1+|m|)(l^{-1}_{c}+|m|a^{-1})}\right)^{2}  
+\hat n k\frac{\partial^{3}h}{\partial x^{3}}  
\end {eqnarray} 
For small slopes, above current generates growth 
equation in the moving frame with average growth rate,  
\begin {eqnarray}
  \frac{\partial h(x,t)}{\partial t}&=&-\frac {F(P_{B}-P_{A})l_{d}}{2}
\frac{\partial^{2} h}{\partial x^{2}}\nonumber \\ 
&&+\frac {Fl^{2}_{d}}{4}   
\frac{\partial^{2}}{\partial x^{2}}\left(\frac{\partial h}{\partial x}\right)
^{2}   
-k\frac{\partial^{4}h}{\partial x^{4}} +\eta(x,t)  
\end {eqnarray} 
Where, $\eta(x,t)$ is the Gaussian noise in the deposition with the 
property, $<\eta(x',t')\eta(x,t)>=\delta(x'-x)\delta(t'-t)$.  
For $P_{A}=P_{B}$, current is tilt free, corresponding equation has the 
Lai- Das Sarma - Villain \cite{ld,vill1} form. 
 
Now consider a top terrace. By definition, only tilt independent terms 
will contribute. Also since steps are absent the only term that contributes 
is, $k\frac{\partial^{3}h}{\partial x^{3}}$ due to the downward hops 
at the edges. For a base terrace, only in-plane hops are possible, hence 
none of the above terms contribute. This shows that these regions 
offer restricted kinetics, thereby changing the growth equation locally. 
If these regions are smaller than $ l_{d}$, they will act only as the regions 
of discontinuity.  
 Villain \cite{vill2} encountered this 
difference in connection with simulations using Zeno equations. However 
so far it has been assumed that a given growth equation is valid over 
the complete substrate and such a distinction is irrelevent under coarsening. 
Based on above argument we propose that {\it stepped, top and base terraces 
are distinct regions} where different growth equation applies. Thus, in 
growth from vapor at low temperature, scalability breaks down. However, 
in simulations and experiments, kinetically rough surfaces are observed.
This is possible provided steps develop in base and top regions at a rate 
comparable with the growth of correlation length defined by the equation over 
steps, shifting these regions. It may be noted that since downward hops 
allow addition of material to lower layers, average shifting of top or base 
region is possible {\it only if }downward hops are present. An infinite SE 
barrier leads to such an immobilization of top and bases\cite{cak}. This is 
the limiting example displaying the effect of three independent regions on   
growth. If 
average shift of top or base regions lags behind the correlation growth, 
mounding instability appears. The lateral growth of the mounds being 
decided by the lateral shift of base regions. It has been established that 
stability of the growth can be quantified in terms of tilt dependent current 
$j_{t}$. For $j_{t}>0$ (uphill), unstable growth while for $j_{t}<0$, 
stable growth is obtained, where $j_{t}$ is measureable in a simulation 
by properly adjusting the boundary conditions\cite{kr1}. Thus, for uphill 
current, base or top regions lag behind the correlations while for 
downhill it shifts at least as fast as correlations. 

Consider unstable growth in 1+1 dimensions. Fig.2 shows the well developed 
mounds. Note the deep ridges formed due to high step heights of the steps 
forming the ridges. The model used for this growth will be described later. 
It suffices to know that its a growth with finite diffusion of adatoms 
and finite SE barrier. We estimate the growth rate by appealing to the 
diffusional kinetics of atoms. The growth proceeds by expansion of a larger 
mound at the cost of smaller one\cite{vill2}. Thus the ridge proceeds in one direction. 
A smaller mound generally poses a smaller angle with respect to the substrate. 
Thus, relatively longer terraces are present on this mound. The diffusional 
addition to the ridges is mainly from these terraces, resulting in to the 
shift of ridge in the direction of smaller mound. Thus, we assume that 
adatoms are added from the smaller mound, diffusionally. The 
diffusional rate of displacement is $dl=D^{1/2}_{s}t^{-1/2}dt$ on a plane 
surface in time $dt$. However, for a ridge to move laterally, it must be 
filled at least up to first step height. For sharp ridges as in Fig.2, the 
step height of ridge may be taken to be $\sim w$, the rms height fluctuation
 (width). Hence, the displacement for a ridge will be $dl_{r}=pD^{1/2}
_{s}t^{-1/2}aw^{-1}$. Where, $p$ is relative fraction of adatoms crossing 
the step edge and $a$ is lattice constant.    
For $w\sim t^{1/2}$, the growth of mounds is 
proportional to $lnt$. 

We verify the $lnt$ dependence for a 1+1 dimensional model that mimics 
the growth at low temperature. In this model, 
on a one dimensional substrate, adatoms are rained randomly. An atom 
with one or more nn is incorporated in the crystal. An adatom with zero nn 
is allowed to hop $n$ number of times at the most. If it acquires a nn, then 
no further hops are allowed. If number of hops are exhausted, it is 
incorporated at the final site after $n$ hops. A parameter $p$ is introduced, 
such that for $p>0.5$ hopping across a step in the downward direction is 
difficult. $p=1$ is the case of infinite SE barrier. We have measured 
$<h_{i}h_{j}>$ correlations for various values of $p$ and used the first 
zero crossing as the measure of the size of the mound. In Fig. 3, plot of 
mound size Vs. time on a semi log scale clearly shows that for $p>0.5$ 
{\it i.e.} for positive SE barrier, the mounds growth is $lnt$.   
Also shown is the case for $p=0.5$.  
We plot length corresponding to the first maximum in height-height 
correlations for this case.  
The curve on semi-log plot is exponential showing a power law dependence. 
Correlation length $\xi\sim t^{1/4}$ in this case. In fact it can be 
shown \cite{prep} that corresponding equation describes 
 Das Sarma - Tamborenea
(DT)\cite{dt} model to which the tilt independent growth equation reduces 
for large slopes. We find that for $p<0.5$, asymptotically, EW growth is 
recovered. Thus, the base and top regions move at least in phase with the 
$\xi$ to provide rough surface. In the present model dissociation from steps
 is not included so that the detailed balance is not followed. 
If this is included, and the current is still uphill 
 then $lnt$ dependence continues for growth in 1+1- dimension.

 Above arguments are true in any dimension. In 2+1- dimensions, mound 
formation is observed experimentally as well as in simulations\cite{len,kr2}. 
Various predictions are referred in the introduction above regarding the 
time evolution of the mounds. The $lnt$ dependence in 1+1- dimensions is 
the upper limit for lateral development of the mounds in 2+1- dimensions. 
This is so because, a given mound is surrounded by four or more mounds. 
Probability that such a mound happens to be the smallest amongst the 
surrounding ones including itself is very small. A given mound may be reduced 
in one direction, but it may increase in other direction owing to a smaller 
 mound there. Thus, instead of consumption, shift of mounds is 
more likely on a two dimensional substrate. In order to find the time 
dependence of mound growth in 2+1- dimensions, we have used same model 
 described above, except that the rules apply in two directions on a 
square lattice. In addition, we have included edge diffusion with {\it no 
edge barriers}. It is observed that edge diffusion suffices to induce 
uphill current so that even if the diffusion of single adatoms is unbiased, 
mound formation is observed. In the absence of edge diffusion but with 
unbiased single adatom diffusion, EW type growth is obtained\cite{pun}.
Noise reduction technique\cite{wl}is employed with reduction factor of 5.  
The growth of mound size is monitored in the same way as for the 1+1- 
dimensions, using zero crossing for the correlations $<h_{i}h_{j}>$.  
Fig.4 shows the plot of mound size as a function of time on semi-log plot. 
Clearly, after an initial growth like $lnt$, the curve tends to saturation, 
confirming the slower growth rate. By varying parameter $p$, a condition 
close to tilt independent current is obtained. The growth in that case 
follows, $t^{1/4}$ power law. From the arguments leading to Eq.4 
, in 2+1- dimensions, we find that asymmetric term will be ineffective if 
step edge tension is lower so that steps morphology is wavy or fingered. 
This is so because, the terrace size can be reduced by step movements 
in the orthogonal directions as well. Thus only $\nabla^{4}h$ term 
contributes, leading to $\beta=1/4$ and $z=4$ in 2+1- dimensions. Clearly, 
this observation suggests that in experimental growth, if SE barrier is 
very small(but nonzero), at low temperature growth rate of mounds can  
be $t^{1/4}$ in the transient region. If the edge tension is high so that 
steps are straight and 
less wavy, asymmetric term can contribute with $\beta$ and $z$, characteristics 
 of a Lai- Das Sarma like equation\cite{ld} in the transition region.

In conclusion, we have shown that growth from vapor on surface proceeds via  
in principle a heterogeneous dynamics. The stepped, base and top regions 
on the surface allow different growth dynamics. As a result the spatial 
scalability breaks down. The effect is distinctly 
observable for unstable growth leading to mound formation. The kinetics 
across the mounds suggest a $lnt$ dependence in 1+1- dimensions which is 
verifiable in a suitable model. A slower growth is predicted in 2+1- dimensions  which is also observed in a model simulation.

\begin{figure} 
\epsfxsize=\hsize \epsfysize = 1.0 in
\centerline{\epsfbox{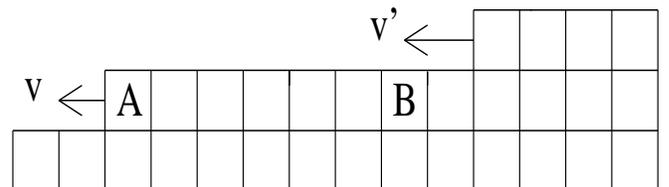}}
\caption{A typical step structure formed during 
growth along positive slope. $v$ and $v'$ are velocities of the steps 
.  
}     
\label{step}  
\end{figure}

\begin{figure} 
\epsfxsize=\hsize \epsfysize= 3.0 in
\centerline{\epsfbox{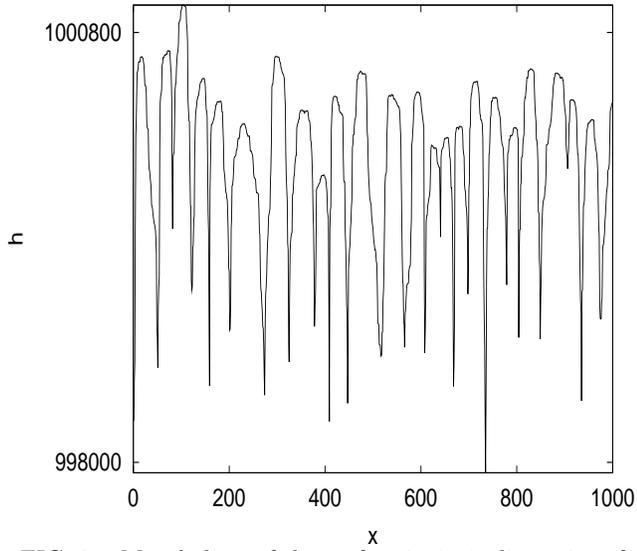}}
\caption{ Morphology of the surface in 1+1- dimensions for an unstable growth  
after $10^{6}$ layers. Parameter $p$ is 0.6.  
}    
\label{morf}  
\end{figure}
 
\begin{figure} 
\epsfxsize=\hsize \epsfysize= 3.0 in
\centerline{\epsfbox{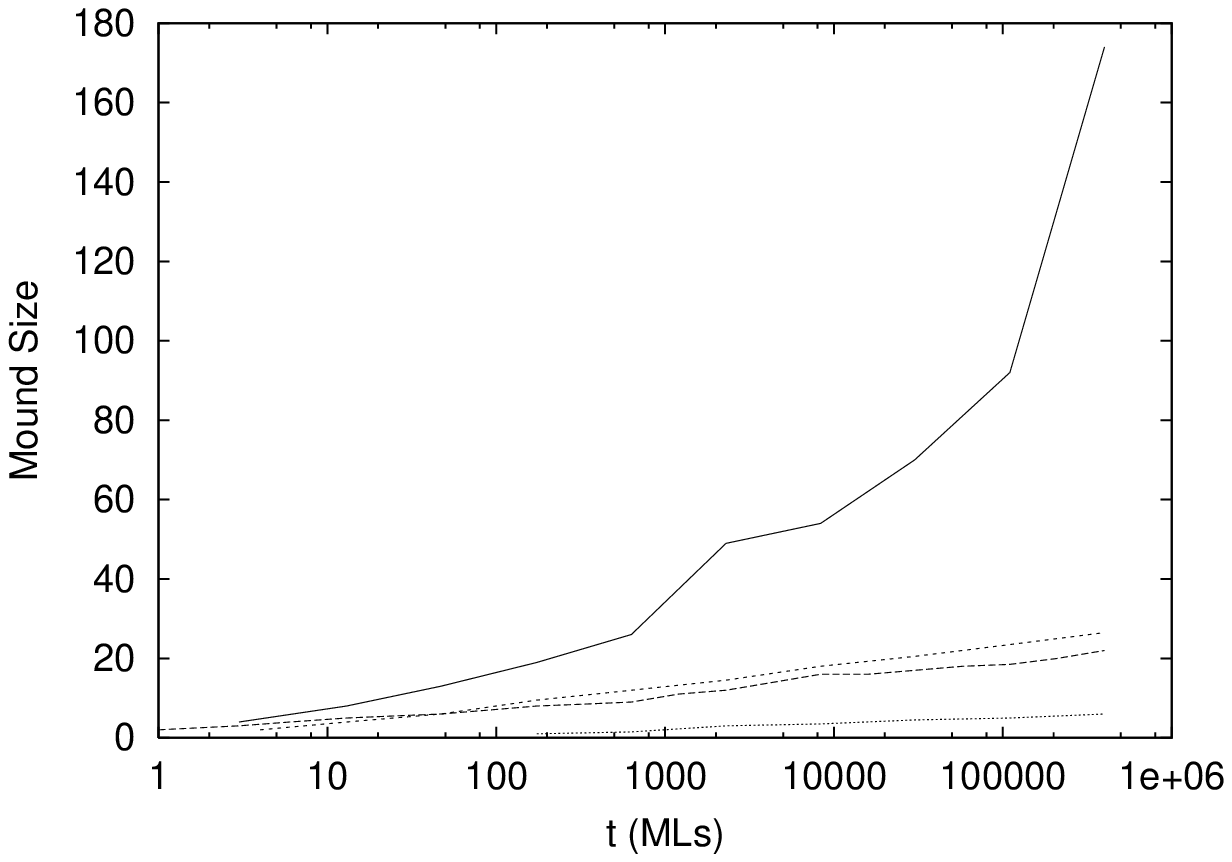}}
\caption{ Shows time evolution of lateral growth in 1+1- dimensions.   
The values of parameter $p$ are 0.5, 0.6, 0.7 and 0.8 respectively for 
the curves from top to bottom in the figure. The substate size is 
$L=10000$.   
}    
\label{cor1}  
\end{figure}

\begin{figure} 
\epsfxsize=\hsize \epsfysize= 3.0 in
\centerline{\epsfbox{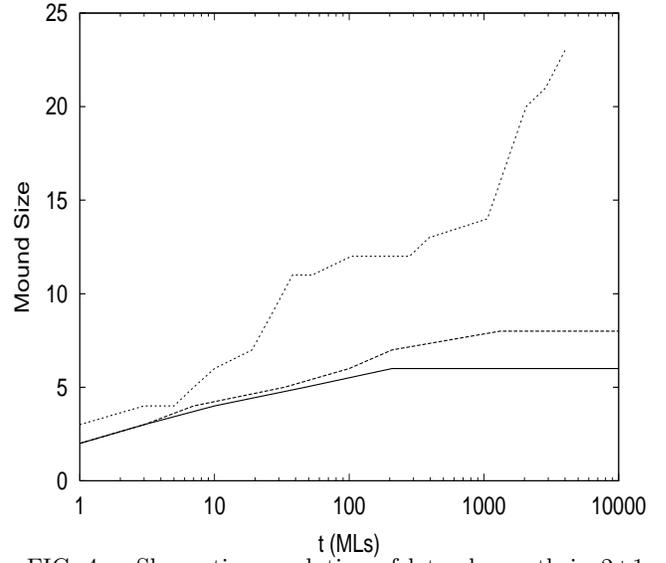}}
\caption{ Shows time evolution of lateral growth in 2+1- dimensions.   
The values of parameter $p$ are 0.35, 0.6, and 0.7 respectively for 
the curves from top to bottom in the figure. The substrate size is 
300 X 300 for the simulation.     
}    
\label{cor2}  
\end{figure}

\end{document}